\documentclass{article}

\usepackage{xurl}
\usepackage{arxiv}
\usepackage{float}
\usepackage[utf8]{inputenc} 
\usepackage[T1]{fontenc}    
\usepackage{hyperref}       
\usepackage{xurl}            
\usepackage{booktabs}       
\usepackage{amsfonts}       
\usepackage{nicefrac}       
\usepackage{microtype}      
\usepackage{lipsum}		
\usepackage{longtable}
\usepackage{graphicx}
\usepackage{natbib}
\usepackage{doi}
\usepackage{multicol}
\usepackage{balance}
\usepackage{array}
\usepackage{multirow}
\usepackage{amsmath}
\usepackage{amssymb}
\usepackage{listings}
\usepackage{xcolor}

\definecolor{codegreen}{rgb}{0,0.6,0}
\definecolor{codegray}{rgb}{0.5,0.5,0.5}
\definecolor{codepurple}{rgb}{0.58,0,0.82}
\definecolor{backcolour}{rgb}{0.95,0.95,0.92}

\lstdefinelanguage{json}{
    basicstyle=\small\ttfamily,
    showstringspaces=false,
    breaklines=true,
    backgroundcolor=\color{backcolour},
    commentstyle=\color{codegreen},
    keywordstyle=\color{magenta},
    stringstyle=\color{codepurple},
}

\title{Making Sense of AI Limitations: How Individual Perceptions Shape Organizational Readiness for AI Adoption}

\author{ {\hspace{1mm}Thomas Übellacker}\thanks{Maastricht University; based on\\ original Master thesis.}
	\texttt{} 
	} 

\renewcommand{\shorttitle}{}

\hypersetup{
pdftitle={MAKING SENSE OF AI LIMITATIONS: HOW INDIVIDUAL
PERCEPTIONS SHAPE ORGANIZATIONAL READINESS FOR AI
ADOPTION},
pdfsubject={q-bio.NC, q-bio.QM},
pdfauthor={David S.~Hippocampus, Elias D.~Striatum},
pdfkeywords={First keyword, Second keyword, More},
}

\begin{document}
\maketitle
\begin{multicols}{2}
\begin{abstract}
	This study investigates how individuals' perceptions of artificial intelligence (AI) limitations influence organizational readiness for AI adoption. Through semi-structured interviews with seven AI implementation experts, analyzed using the Gioia methodology, the research reveals that organizational readiness emerges through dynamic interactions between individual sensemaking, social learning, and formal integration processes. The findings demonstrate that hands-on experience with AI limitations leads to more realistic expectations and increased trust, mainly when supported by peer networks and champion systems. Organizations that successfully translate these individual and collective insights into formal governance structures achieve more sustainable AI adoption. The study advances theory by showing how organizational readiness for AI adoption evolves through continuous cycles of individual understanding, social learning, and organizational adaptation. These insights suggest that organizations should approach AI adoption not as a one-time implementation but as an ongoing strategic learning process that balances innovation with practical constraints. The research contributes to organizational readiness theory and practice by illuminating how micro-level perceptions and experiences shape macro-level adoption outcomes.
\end{abstract}


\section{Introduction}

\vspace{-3mm}
Artificial Intelligence (AI) has emerged as a new technology platform, changing how organizations operate and compete. The rapid advancement of AI capabilities, particularly in areas such as Large Language Models (LLMs) and multimodal systems, has created new opportunities for organizations to automate complex tasks, enhance decision-making processes, and drive innovation \citep{Bommasani2022,Bubeck2023,Li2023}. Organizations across industries are increasingly seeing AI adoption as a strategic imperative, with 55\% of organizations reporting the use of AI in at least one business unit or function as of 2023, up from 50\% in 2022 \citep{Maslej2024} and with global AI spending projected to reach \$631 billion by 2028 \citep{Massey2024}. However, organizations struggle to implement AI despite investments and evident strategic importance. They encounter challenges beyond technical considerations, encompassing social and organizational dynamics. AI adoption initiatives often fail to deliver their intended benefits, with reasons such as unrealistic expectations, lack of change management, organizational constraints, organizational readiness, and failure to understand users' needs identified as barriers to successful adoption \citep{Cooper2024,Westenberger2022}.

Understanding organizational readiness for technological change has become increasingly critical as organizations face the challenges of AI adoption. While traditional models of organizational readiness have focused on structural and technical preparedness (e.g., \cite{Weiner2009}), the unique characteristics of AI technologies - their complexity, opacity, and transformative potential - demand a more nuanced understanding of how organizations become ready for AI adoption. The role of individual perceptions in forming this readiness is noteworthy, as individuals' understanding and interpretation of AI's limitations can influence an organization's capacity to implement AI systems successfully. These perceptions are not formed in isolation but are shaped through complex social and organizational processes as individuals attempt to make sense of AI's capabilities, limitations, and implications for their work.

Sensemaking theory \citep{Weick2005} provides a valuable theoretical lens for understanding how individuals interpret and make meaning of AI technologies and their limitations. Through sensemaking processes, individuals construct their understanding of what AI technologies are, how they function, and what implications they have for their work and organizational practices. These interpretations, in turn, shape organizational readiness for AI adoption. However, while existing research has broadly examined organizational readiness for technological change \citep{Armenakis2002,Holt2007}, a theoretical gap exists in understanding how individuals' perceptions of AI-specific limitations influence organizational readiness for AI adoption. Despite the growing body of research on AI implementation and its challenges, we lack a theoretical framework that explains how individuals' sensemaking of AI limitations shapes organizational readiness for AI adoption.

This study seeks to address this gap by examining the following research question: \textit{How do individuals' perceptions of AI limitations influence organizational readiness for adopting AI technologies}? By exploring the interpretive processes through which individuals construct their perceptions of AI limitations and how these perceptions influence organizational readiness, this study seeks to develop a more comprehensive understanding of the social and psychological factors that shape AI adoption success.

This study contributes to organizational readiness theory by showing how employees' firsthand experiences of AI limitations - from biases to technical constraints - eventually shape an organization's preparedness for AI adoption. By showing that individual perceptions are enablers or impediments to broader organizational adoption, the research highlights the role of social and experiential processes in driving adoption outcomes. Further, it enriches sensemaking theory by revealing how encounters with AI limitations trigger collective interpretation and trust-building processes, which influence organizational readiness. These insights result in a novel theoretical framework connecting individual-level sensemaking with macro-level readiness dynamics, offering a more holistic view of AI adoption.

In practical terms, the findings give managers actionable strategies to guide AI initiatives effectively. First, they demonstrate how organizations can exploit hands-on experimentation and peer learning to interpret AI limitations to allow user expectations to remain realistic. Second, they show how systematic governance, clear policies, and purposeful integration into existing workflows help solidify organizational readiness. Finally, the research encourages developing an environment that supports iterative learning and trust development, enabling a climate where employees openly engage with AI's evolving capabilities and limitations. These recommendations would allow managers to address individual concerns, refine change management processes, and build an organizational culture that positions AI as a sustainable and value-increasing resource.

\section{Literature Review}

The relationship between individual perceptions of AI limitations and organizational readiness is a multi-level phenomenon that cannot be fully understood through traditional technology adoption frameworks alone. An integrated theoretical approach incorporating sensemaking processes trust development mechanisms and organizational readiness dynamics is needed.

This literature review approaches AI adoption and readiness as a multi-level phenomenon. It addresses external pressures (such as competitive forces, policy frameworks, and societal discourse), organizational-level readiness (in terms of capabilities, culture, and infrastructure), and finally, individual-level factors (such as perceptions of AI limitations and sensemaking processes). The interplay across these levels highlights how organizations successfully adopt AI.

\subsection{AI Adoption in Organizations}

AI adoption has several unique characteristics that differentiate it from traditional technology adoption. \cite{Weber2023} identify two characteristics: inscrutability and data dependency. Inscrutability manifests in the difficulty of predicting system behavior and explaining decision processes, while data dependency requires continuous system adjustments as organizational data evolves. These characteristics create increased variability in organizational decision-making processes that require new coordination mechanisms \citep{Agrawal2024}. The inscrutability concept is particularly interesting for understanding individual interactions with AI systems, as it directly influences how organizational members interpret and respond to AI-driven changes.

\cite{Agrawal2024} argue that AI adoption increases decision variation across interconnected organizational tasks, asking organizations to either reduce task interdependencies or implement strong coordination mechanisms. This finding challenges the typical focus on individual task-level AI adoption by highlighting the systemic nature of organizational AI adoption. \cite{Yang2024} extend this understanding by showing how AI adoption introduces both technological affordances and constraints that vary significantly based on organizational size. Their study reveals that while larger firms perceive primarily operational affordances focused on efficiency and quality improvements, smaller firms see marketing affordances as more important, leading to different adoption patterns and outcomes.

With his Technology-Organization-Environment (TOE) framework, \cite{Baker2012} emphasizes that innovation adoption depends on the interplay between technological features, organizational characteristics, and environmental conditions. Applied to AI adoption, \cite{Yang2024} show how these elements are expressed through innovation management approaches, organizational AI readiness, and environmental pressures. This can also serve as a framework for looking at adoption barriers. \cite{Cubric2020} categorizes them across technical aspects (data availability, model reusability), organizational considerations (resource allocation, support infrastructure), and social dimensions (human-AI interaction, job security concerns, trust issues).

The organizational context shapes adoption patterns through what \cite{Weber2023} identify as implementation capabilities. These capabilities encompass AI project planning, co-development, data management systems, and model lifecycle management processes. \cite{Alekseeva2020} demonstrate that AI adoption among management ranks, rather than just IT specialists, drives positive organizational outcomes. This finding highlights the importance of broad organizational involvement in the adoption process and suggests that successful AI implementation requires capabilities across different organizational levels.

Adopting AI technologies follows patterns that reflect unique characteristics and organizational implications. \cite{Henry2022} emphasize the importance of human-machine teaming in successful AI adoption, noting that individuals build trust with AI systems through experience, expert endorsement, and systems designed to accommodate professional autonomy. This observation aligns with \cite{Kelley2022} identification of success factors, including effective communication channels, management support, training, and established reporting mechanisms for addressing AI-related concerns.

These implementation patterns reveal several sensitizing concepts important for understanding AI adoption. First, AI inscrutability is fundamental in shaping how organizational members interact with and interpret AI systems. Second, the system-wide impact concept captures the extensive organizational changes that AI adoption necessitates. Third, adoption barriers appear across technical, organizational, and social dimensions, providing observable indicators of adoption challenges. Fourth, capability distribution reflects the spread of AI-related competencies across organizational levels. Finally, implementation patterns capture organizations' observable approaches to managing AI adoption.

The complexity and uniqueness of AI adoption create distinct challenges requiring careful consideration of technical and organizational dimensions. These challenges are particularly evident in what \cite{Cubric2020} identifies as the social considerations of AI adoption, including increased dependence on non-human agents, job security concerns, and trust issues. Understanding these dynamics provides context for examining how organizations develop readiness for AI adoption, particularly considering the role of individual perceptions in shaping adoption outcomes.

Beyond these internal adoption challenges, external pressures - such as industry competition, global technology trends, and regulatory policy - further shape the path to AI adoption \citep{Yang2024}. \cite{Felemban2024} highlight the role of government support in the Saudi context, showing how initiatives like Vision 2030 influence adoption through multiple channels: directly through regulatory frameworks and policies and indirectly by shaping senior management support and competitive dynamics between organizations. Their study reveals that government support affects all aspects of the technology-organization-environment framework, creating opportunities and pressures for adopting organizational AI.

\subsection{Organizational Readiness for AI Adoption}

Organizational readiness is not only about having the right resources or leadership in place; it also mediates between broad external pressures (e.g., policy mandates and competitive landscapes) and how employees on the ground perceive and engage with AI. Organizational readiness serves as a bridge between broader external forces (such as policy requirements and market competition) and how individual employees actually engage with and implement AI in their daily work. It determines how well an organization can translate high-level strategic demands into successful adoption by its workforce. \cite{Johnk2021} emphasize that readiness involves aligning organizational assets, individual capabilities, and leadership commitment to support AI initiatives. They identify five core domains – strategic alignment, resources, knowledge, culture, and data - that collectively determine readiness.

AI readiness demands more than just technical infrastructure. \cite{Heimberger2024} highlight that success depends on how well organizational processes can integrate AI. That includes adapting workflows, ensuring data compatibility, and developing continuous learning and refinement systems. Readiness is, therefore, an evolving state influenced by the organization's ability to adjust and respond to AI's changing demands—a classical organizational learning problem.

Organizations need to develop specific capabilities for successful AI adoption. \cite{Weber2023} identified four concrete organizational capabilities: AI project planning, co-development of AI systems, data management, and AI model lifecycle management. This is a more process-oriented approach to readiness than the readiness factors \cite{Johnk2021} synthesized.

Leadership is vital for AI readiness. \cite{Felemban2024} argue that senior management support significantly affects individual attitudes and readiness to adopt AI. Leaders are important in allocating resources, prioritizing AI in strategic plans, and addressing resistance from change recipients \citep{MikelHong2024}.

Trust in AI systems is another determinant for organizational readiness, influencing adoption and sustained engagement \citep{Tursunbayeva2024}. \cite{Glikson2020} emphasize that trust involves cognitive elements, like reliability and transparency, and emotional elements, such as the perceived human-likeness of AI. \cite{Thiebes2021} highlight trustworthiness principles - beneficence, non-maleficence, autonomy, justice, and explicability - as critical to fostering trust. Building trust requires consistent system reliability, ethical alignment, and clear explanations of AI behavior. \cite{Siau2018} emphasize the importance of expert endorsements, validation, and iterative user interactions in increasing trust. Similarly, \cite{Henry2022} found that integrating AI into workflows while respecting user autonomy strengthens trust by framing AI as a collaborative tool rather than a replacement. Addressing these dimensions ensures individuals view AI as reliable and aligned with their roles, reducing resistance and enabling successful adoption.

Individual employees' cultural values collectively shape another critical dimension of organizational preparedness for AI adoption, as these personal orientations aggregate to influence the organization's overall cultural readiness for technological change. According to \cite{Sunny2019}, individual cultural values significantly impact technology acceptance and readiness. Their research found that collectivism and long-term orientation positively influence the perceived usefulness and ease of use of new technologies at the individual level. Additionally, they found that a less masculine organizational culture helps reduce employee discomfort with technological change. \cite{Hradecky2022} find that organizations, particularly in the exhibition industry, struggle with cultural barriers, such as risk aversion and resistance to change, which hinder readiness. Conversely, a culture of openness and collaboration can drive more effective adoption processes.

Organizational readiness is not developed in isolation but interacts with external pressures and opportunities. \cite{Yang2024} highlight how competitive environments drive organizations to develop AI capabilities aggressively. Further, government and regulatory support play a significant role in shaping readiness. Indeed, policy-level initiatives can catalyze AI readiness by providing resources or mandating standards \citep{Felemban2024}.

Although organizational readiness lays the strategic and cultural groundwork for AI adoption, its success ultimately depends on how individual employees perceive and integrate these technologies into their work. \cite{Johnk2021} highlight the role of workforce capabilities, particularly in developing skills and trust in AI systems. Individuals who view AI as threatening their autonomy or job security may resist its implementation. Addressing these concerns through communication, training, and involvement in AI projects can increase readiness. Hence, the following section turns to the micro-level factors that can accommodate or undermine readiness.

\subsection{Individual Perceptions of Limitations}

The successful adoption of AI technologies within organizations is not solely determined by technical capabilities but is significantly influenced by individual perceptions of AI limitations \citep{Glikson2020,Kelley2022}. These perceptions can act as barriers or facilitators to adoption, affecting organizational readiness and the overall implementation process \citep{Trenerry2021}.

Individuals' perceptions of AI limitations encompass a range of concerns that can hinder the adoption of AI technologies within organizations. One area of concern is the issue of trust and reliability. \cite{Nasarian2024} and \cite{Xiangwei2022} highlight that individuals find building trust in AI systems difficult due to inconsistent or opaque outputs. This lack of trust is further impeded when AI systems fail to perform reliably in critical applications, leading to skepticism about their usefulness, as \cite{Singh2023} noted. Additionally, \cite{Choudhary2024} observe that fear and resistance to adoption can come from a misalignment between AI technologies and individuals' values and anxiety over dealing with complex IT systems.

Transparency and explainability of AI systems are also significant concerns among individuals. The "black box" nature of AI algorithms, particularly in complex models like large language models (LLMs), poses challenges for those who require clear and interpretable decision-making processes. \cite{Novak2022} and \cite{Haxvig2024} discuss how the lack of transparency can hinder individuals' understanding and acceptance of AI outputs. \cite{Lai2023} and \cite{Morais2023} further point out that AI systems often cannot provide meaningful, user-aligned explanations, which can decrease trust and confidence among users.

Furthermore, concerns about human-AI interaction play a role in shaping individuals' perceptions. \cite{Qian2024} and \cite{Bucinca2021} note that individuals may be cautious of over-reliance on AI and the potential for automation complacency, leading to skill degradation or reduced caution in their roles. Additionally, cognitive and self-serving biases can influence how individuals interpret AI capabilities. \cite{vonSchenk2023} demonstrate that when people lack information about how AI systems operate - specifically about what happens to machines' earnings in economic interactions - they tend to form self-serving beliefs that justify less cooperative behavior with the machines. The lack of emotional intelligence in AI systems, especially in contexts requiring empathy and nuanced human interaction, is another limitation that \cite{Singh2023} cited.

Bias and fairness issues embedded in AI systems are significant concerns that affect individuals' willingness to adopt these technologies. \cite{Muller2022} discuss how biases in training data can affect AI outputs, leading to unfair or discriminatory outcomes. \cite{Allan2024} and \cite{zhou2023teaching} emphasize that amplifying societal stereotypes through AI systems poses ethical and legal risks, prompting individuals to question the fairness and appropriateness of AI-driven decisions within their organizations.

Technical limitations, such as inconsistent performance, contribute to individuals' skepticism about AI technologies. \cite{Muller2022} and \cite{Lin2024} report that perceived inconsistencies in AI performance can undermine individual confidence in these systems. \cite{Pinto2024} and \cite{Biswas2023} highlight that AI systems' difficulties in processing nuanced or context-specific information relevant to specific tasks can further diminish individuals' perceptions of AI effectiveness.

Ethical considerations also shape individuals' perceptions of AI limitations. \cite{Whittle2021} and \cite{Aliman2021} note that potential societal harm due to bias, misinformation, or unethical use can lead to resistance among individuals prioritizing ethical standards in their work. \cite{zhang2024symlearn} and \cite{Fang2023} observe that gaps between ethical principles and their implementation in AI technologies can result in individuals questioning the adoption of such systems.

Finally, practical challenges in implementing and integrating AI systems into existing workflows are perceived as significant limitations. \cite{Boukhelifa2020} identify key challenges in interactive AI systems, including difficulties in defining appropriate roles between humans and AI, managing trade-offs between competing objectives like accuracy and interpretability, and dealing with multiple sources of uncertainty. The challenges of integrating AI can also vary by context - for instance, in academic writing, \cite{Chemaya2024} found disagreement among academics about appropriate AI use and reporting requirements, with differences shaped by role, ethics perceptions, and language background. \cite{zhang2023human} found that while there was no broad aversion to AI systems, persistent human favoritism could affect integration efforts.

These perceived limitations align with key sensitizing concepts such as AI inscrutability and the adoption barriers across technical, organizational, and social dimensions discussed by \cite{Weber2023}, \cite{Cubric2020}, and \cite{Agrawal2024}. Understanding these perceptions is important for organizations aiming to improve their readiness for AI adoption, as they influence individuals' willingness to engage with and support the integration of AI technologies.

According to \cite{Moore1991}, perceptions of adopting an information technology innovation are shaped by factors such as relative advantage, compatibility, complexity, trialability, and observability. In AI adoption, individuals' prior experiences with technology, individual innovativeness, and organizational communication channels contribute to perception formation \citep{Agarwal1998,Haenssgen2018}. For instance, individuals with higher personal innovativeness are more likely to develop positive perceptions of AI technologies \cite{Agarwal1998}.

The domain of uncertainty framework suggests that uncertainties associated with change fit into four domains: conceptual uncertainty (What is the change?), functional value uncertainty (What is the value of the change?), process uncertainty (How will the change come about?), and impact uncertainty (What is the broader impact of the change?) \citep{Yin2024}. Individuals form perceptions based on how AI technologies address these uncertainties. For example, conceptual uncertainty arises from a lack of understanding of AI's functionalities, while impact uncertainty pertains to doubts about AI's long-term effects on job security and organizational practices.

External factors such as media representations, societal discourse, and organizational communication strategies influence perception formation \citep{Agarwal1998,Trenerry2021}. The perceived risks and uncertainties associated with AI, including job displacement and ethical concerns, are amplified or mitigated through these channels \citep{FakhrHosseini2024,Sadeck2022}. Communication channels shape perceptions, as individuals rely on mass media and interpersonal communications to develop their understanding of AI technologies \citep{Agarwal1998}.

Finally, the literature highlights that individuals' perceptions of AI limitations are shaped through interpretation and meaning-making processes. Individuals attempt to understand how AI fits into their professional roles, organizational goals, and broader societal contexts, reconciling uncertainties about transparency, fairness, and ethical alignment \citep{Maitlis2014,Yin2024}. This interpretive process is both individual and collective, as organizational culture, peer interactions, and shared assumptions influence how individuals construct their understanding of AI technologies. \cite{Orlikowski1994} introduce the concept of "technological frames," highlighting how shared assumptions and knowledge within organizations shape people's perceptions and interactions with technology. Similarly, \cite{Balogun2005} demonstrate how informal networks and lateral employee interactions contribute to evolving interpretations during organizational change. These dynamics suggest that perceptions of AI are formed through ongoing collective processes at both personal and organizational levels. Understanding these shared interpretations offers valuable insights into how readiness and adoption are shaped, which will be examined in greater depth in the following section.

\vspace{-3mm}
\subsection{Sensemaking}

Sensemaking theory provides a valuable framework for understanding how individuals, teams, and organizations interpret and respond to AI. This approach is inherently multi-level, encompassing the personal sensemaking of employees, the collective sensemaking of groups or departments, and organizational sensemaking processes \citep{Maitlis2014,Weick2005}. These nested sensemaking processes also incorporate external cues - such as media stories, industry regulations, and competitive forces - reinforcing that AI adoption is shaped by influences from the macro-level to the micro-level.

Sensemaking is triggered by cues that disrupt individuals' existing understanding, prompting them to seek explanations and restore meaning \citep{Maitlis2014}. With their unique characteristics and potential implications, introducing AI technologies can serve as such a trigger, creating a need for individuals to make sense of these new realities \citep{Yin2024}. \cite{Weick1995} identifies seven properties of sensemaking: identity construction, retrospection, enactment, social interaction, ongoing nature, extraction of cues, and plausibility over accuracy. These properties provide a framework for examining how individuals interpret AI limitations and construct their understanding of the technology's role in their work. Identity construction is central to sensemaking, as individuals interpret events in ways that maintain a consistent self-conception \citep{Weick1995}. In the context of AI adoption, individuals may perceive AI limitations in ways that align with their professional identities and values \citep{Choudhary2024}.

Sensemaking is inherently retrospective, as individuals make sense of events by drawing on past experiences and existing frameworks (Weick, 1995). Individuals' prior experiences with technology adoption and their exposure to the societal discourse around AI can shape their interpretations of AI limitations \citep{Agarwal1998,Trenerry2021}. This retrospective nature also suggests that individuals' perceptions may evolve as they accumulate experiences with AI technologies over time.

Enactment is another property of sensemaking, emphasizing that individuals actively construct the environments they face \citep{Weick1995}. In the context of AI adoption, individuals' actions and responses to the technology can shape the organizational reality surrounding AI. For instance, following what we know about confirmation bias, resistance, or avoidance behaviors based on perceived limitations could create self-fulfilling prophecies, reinforcing the challenges of AI integration \citep{Peters2022}.

Social interaction is important to sensemaking, as individuals rely on shared narratives and collective interpretations to construct meaning \citep{Weick1995}. Individuals' perceptions of AI limitations are not formed in isolation but are influenced by interactions with colleagues, organizational communication, and broader societal discourse \citep{Agarwal1998,Trenerry2021}. The social nature of sensemaking suggests that organizations can actively shape individuals' perceptions through strategic communication and promoting a supportive culture around AI adoption.

Sensemaking is an ongoing process, as individuals continuously update their interpretations based on new information and experiences \citep{Weick1995}. This ongoing nature is particularly relevant in the rapidly evolving AI landscape, where individuals' perceptions may shift as they encounter new applications, capabilities, and challenges. Organizations need to recognize the dynamic nature of sensemaking and provide ongoing support and communication to help individuals navigate the evolving realities of AI adoption.
The extraction of cues refers to the process by which individuals selectively attend to certain aspects of their environment to support their interpretations \citep{Peters2022,Weick1995}. In the context of AI adoption, individuals may focus on cues that reinforce their existing perceptions of AI limitations, such as instances of biased outputs or technical failures. Change agents could actively manage the cues available to individuals by highlighting successful AI implementations and providing transparent information about the technology's capabilities and limitations.

Finally, sensemaking prioritizes plausibility over accuracy, as individuals seek interpretations that are sufficiently coherent and credible to guide action \citep{Weick1995}. Individuals' perceptions of AI limitations may not always align with the technology's objective realities but are constructed in ways that make sense given their experiences, beliefs, and organizational context. This suggests that organizations must create narratives and experiences that promote positive and plausible interpretations of AI's role in the workplace.

Sensemaking perspective aligns with key insights from the previously discussed literature, such as the importance of addressing conceptual and impact uncertainties \citep{Yin2024}, the role of communication channels in shaping perceptions \cite{Agarwal1998}, and the influence of organizational culture on adoption readiness \cite{Johnk2021}. Sensemaking theory extends these insights by providing a framework for understanding the cognitive and social processes through which individuals actively construct their perceptions of AI limitations.

Moreover, a multi-level sensemaking theory offers a dynamic and process-oriented view of perception formation, complementing the more static factors emphasized in technology acceptance models like TAM and UTAUT \citep{Davis1989,Venkatesh2003}. By recognizing the ongoing and retrospective nature of sensemaking across different levels (macro to micro), organizations can develop more responsive and adaptive strategies for managing individuals' perceptions throughout the AI adoption process.

However, sensemaking theory also highlights the challenges of managing perceptions in the face of technological complexity and uncertainty. AI technologies' inscrutability and data dependency \citep{Weber2023} can make it difficult for individuals to extract clear cues and construct plausible interpretations. The rapidly evolving capabilities of AI may also require continuous updating of sensemaking frameworks, placing demands on individuals and organizations to remain adaptable.

Bridging the gap between individual-level sensemaking and organizational-level readiness can be understood through organizational learning frameworks, such as the "4I" framework by \cite{Crossan1999}. This framework conceptualizes learning as a multi-level process composed of four stages: intuiting, interpreting, integrating, and institutionalizing. At the individual level, individuals' intuit' and 'interpret' cues derived from their encounters with AI technologies and their perceived limitations. For example, employees may intuitively feel uncertainty or distrust when encountering opaque AI-driven decisions. Through personal interpretation, they construct a narrative that explains why the system behaves unpredictably. These individually held narratives converge as individuals engage in conversations and share experiences, moving from isolated interpretations to more collectively shared meanings.

Once collective interpretations solidify, the process shifts into 'integrating' at the group level. Teams develop a shared understanding of AI's limitations - its inscrutability, data dependencies, or fairness issues - and collectively decide how to respond. Over time, these group-level interpretations become 'institutionalized' into organizational practices, policies, and routines, shaping how the organization prepares for, manages, and leverages AI. Thus, individual sensemaking about AI limitations diffuses upward through group interactions and ultimately informs the organization's formal systems and culture, influencing organizational readiness for AI adoption. In this way, the alignment (or misalignment) between individual interpretations and organizational-level structures and strategies determines how effectively the organization can integrate AI technologies into its core operations.

While the sensemaking perspective provides valuable insights into how individuals interpret and make meaning of AI technologies, the next step is to distill key concepts that can guide the empirical investigation. Drawing from the literature reviewed above, several sensitizing concepts are particularly relevant for understanding how individuals' perceptions of AI limitations influence organizational readiness. These concepts serve not as rigid theoretical constructs but as flexible guides that orient the investigation while remaining open to emergent themes and patterns.

\subsection{Sensitizing Concepts}

The literature suggests several interconnected sensitizing concepts that operate across multiple levels - from individual cognition to organizational processes to external influences. These concepts guide the empirical investigation and anticipate the dynamic relationships that emerge in the findings. The concepts are organized to reflect how perceptions of AI limitations flow from individual interpretation through collective sensemaking to organizational adaptation. These concepts also guide the empirical inquiry into how organizations navigate AI adoption, from external demands and industry-wide influences to individual employees' daily interpretations and actions. The literature on AI adoption, organizational readiness, individual perceptions, and sensemaking suggests several interconnected concepts that inform exploration in further empirical investigation. These sensitizing concepts provide a basis for understanding how individuals' perceptions of AI limitations influence organizational readiness for adoption.

The literature suggests that how individuals form and develop their perceptions of AI limitations is a complex process influenced by individual and contextual factors. Understanding how people identify and categorize different types of limitations is crucial at the individual level \citep{Muller2022,zhang2023human}. Professional background and expertise shape these interpretations, with individuals from different functional areas potentially perceiving limitations differently \citep{Henry2022}.

Contextual influences emerge as equally important in perception formation. The organizational environment, including existing technological infrastructure and support systems, shapes individuals' perceptions of limitations \citep{Weber2023}. Industry-specific challenges and opportunities create unique contexts influencing perception formation \citep{Yang2024}. External discourse, including media representation and professional networks, also contributes to how individuals understand and interpret AI limitations \citep{Trenerry2021}.

The literature highlights sensemaking at both individual and collective levels in how people interpret and respond to AI limitations. At the individual level, sensemaking involves personal interpretation and meaning-making processes \cite{Weick1995}. Individuals engage in retrospective reflection on their experiences with AI, drawing on their professional identity and past experiences to make sense of the limitations they encounter \citep{Maitlis2014}. This individual sensemaking process is ongoing, as people continuously update their interpretations based on new experiences and information.

The collective dimension of sensemaking emerges through social interactions and shared meaning construction. Knowledge sharing is an important mechanism, with groups developing shared understandings through formal and informal discussions \cite{Weick1995}. Informal networks are particularly important in sharing experiences and interpretations across organizational boundaries. These collective processes do not replace individual sensemaking but interact with it, as individuals draw on collective interpretations while contributing their understanding to the group's sensemaking process. Social dynamics within organizations influence both individual and collective sensemaking. Peer experiences and opinions shape interpretations, while leadership is essential in framing how limitations are understood and addressed \citep{Felemban2024}.

The literature suggests that organizational readiness develops through distinct patterns influenced by organizational responses and cultural evolution. Organizations adapt to perceived limitations through various mechanisms, including resource allocation decisions and capability development \citep{Johnk2021}. These responses shape the organization's overall readiness for AI adoption.

Cultural evolution appears to be an important aspect of readiness development. Organizations change work practices and routines as they adapt to AI technologies. Shifts in organizational attitudes and the development of learning processes emerge as important elements of this evolution. Patterns of resistance and acceptance also play a significant role in how readiness develops over time.

These sensitizing concepts suggest several key areas for exploration in this empirical investigation. They emphasize the importance of examining individual experiences and collective processes to understand how perceptions influence readiness. They also highlight the need to consider formal organizational responses and informal social dynamics. The concepts serve as a base for the interview guide (see Appendix 12.1), suggesting areas of inquiry while remaining open to emergent themes. These concepts remain deliberately broad to allow for unexpected findings and emerging patterns during data collection. They orient the investigation while maintaining flexibility to explore new directions as they emerge from the interviews.
\end{multicols}

\begin{figure}[!ht]
    \centering
    {\includegraphics[width=1\linewidth]{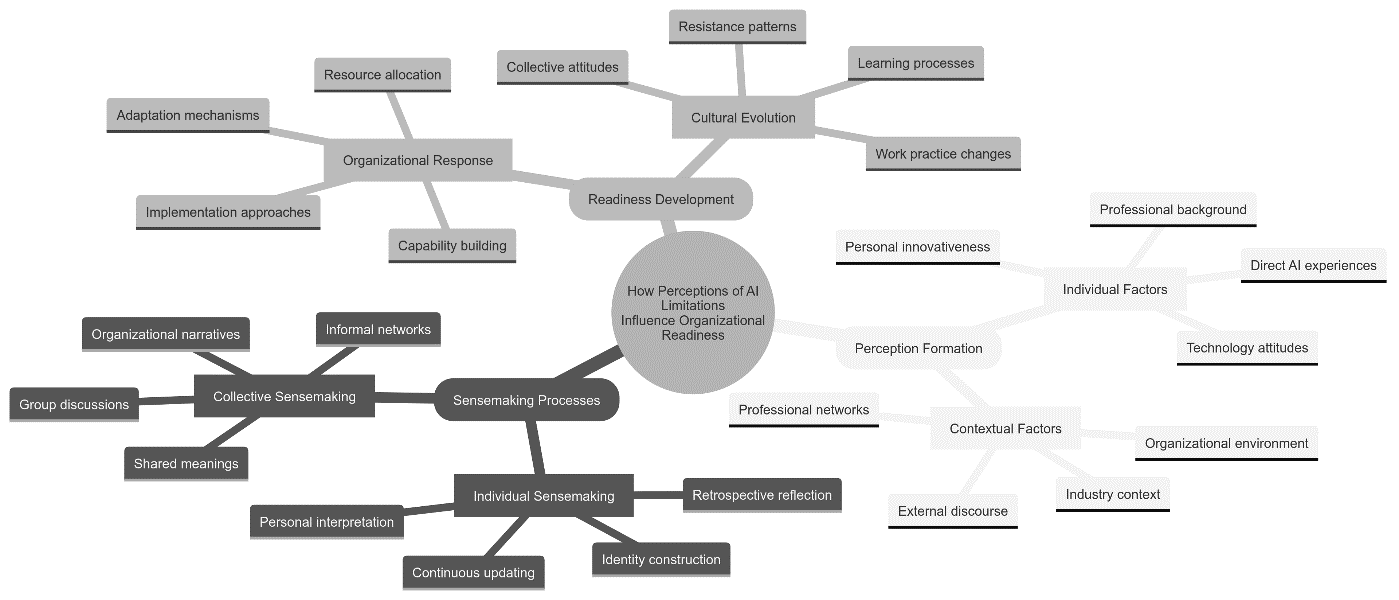}}
    \caption{Sensitizing Concepts}
    \label{fig:fig1}\vspace{10mm}
\end{figure}

\begin{multicols}{2}

\section{Methodology}

\vspace{-3mm}
\subsection{Research Design}
The study employs a qualitative research design grounded in the principles of grounded theory development. Grounded theory allows for developing a theoretical framework that emerges directly from empirical data \citep{Charmaz2012}, making it suitable for investigating how perceptions of AI limitations influence organizational readiness.

\subsection{Epistemological Considerations}
This study is grounded in a constructivist epistemology, which posits that reality is socially constructed through individual and collective interpretations and interactions \citep{Young2004}. Constructivism is appropriate for this research as it emphasizes understanding the subjective meanings that individuals assign to their experiences with AI technologies and how these meanings influence organizational readiness for adoption. By adopting a constructivist lens, the research seeks to co-construct knowledge with participants through in-depth interviews, allowing for a rich exploration of how perceptions of AI limitations emerge and impact organizational readiness. This approach is consistent with qualitative methodologies such as grounded theory and the Gioia method, prioritizing participants' perspectives and the meanings they ascribe to phenomena \cite{Gioia2013}. The constructivist epistemology supports grounded theory in allowing theories to emerge from the data rather than imposing preconceived hypotheses \citep{Mills2006}. This is particularly relevant for exploring new and complex phenomena like AI adoption in organizations, where existing theories may not fully capture the intricacies of human perceptions and social processes.

\subsection{Data Collection}

To comprehensively address the research question, this study employed a qualitative data collection method using semi-structured expert interviews (see interview guide in Appendix 12.1). This methodological choice aligns with the exploratory nature of the research question and its focus on understanding complex organizational phenomena \citep{Eisenhardt1989}. Semi-structured interviews are particularly suited for capturing rich insights about technology adoption processes while maintaining systematic data collection \citep{Gioia2013}.

The participant selection uses a purposive convenience sampling strategy, which proves valuable for accessing experts with rich insights on the studied topic \citep{Etikan2016}. Participants were recruited through two primary channels: personal professional networks (n=3) and the AI Impact Mission community, an active online community of approximately 330 LLM enthusiasts (n=4). This dual-channel approach helped ensure access to participants with deep expertise in AI implementation while maintaining the diversity of perspectives. Informants were selected based on the following criteria: (1) current or recent (within the last two years) involvement in strategic or technical leadership roles; (2) experience with multiple AI adoption projects across organizational contexts; and (3) understanding of both technical and organizational aspects of AI implementation. Selected participants had strategic or technical roles such as "AI Consultant," "Head of AI," and "AI Systems Engineer."

The sampling strategy prioritized information richness over representativeness, focusing on participants who could provide deep insights into AI adoption processes based on their direct involvement in implementation projects across various organizational contexts. This approach aligns with qualitative research best practices, emphasizing depth and quality of insights rather than statistical generalizability \citep{Patton2002}.

The final sample consisted of seven participants, though additional potential participants had expressed interest in participating. The decision to conclude data collection at seven interviews was guided by data saturation \citep{Guest2006}, which was systematically assessed through quantitative analysis of new code generation (see Appendix 12.2). The analysis revealed a clear pattern of diminishing returns in terms of new insights generated from each subsequent interview. The first interview yielded 105 unique codes, establishing the initial conceptual framework. The second interview contributed 83 new codes, expanding the theoretical understanding significantly while validating many concepts from the first interview. A drop in new code generation was observed with the third interview, which added 52 new codes, suggesting the beginning of saturation.

The pattern of diminishing returns became more pronounced in subsequent interviews, with interviews four through seven, each contributing between 11-25 new codes (see Figure \ref{fig:fig1}). While these later interviews provided valuable validation and a nuanced understanding of existing concepts, the limited number of new codes suggested that the core theoretical categories had been well-established. The cumulative number of unique codes reached approximately 320, with the curve showing clear signs of plateauing. This plateauing effect and consistent validation of themes in later interviews provided strong evidence that theoretical saturation had been achieved.

Interviews were conducted remotely via Microsoft Teams over seven days, lasting between 40 and 60 minutes. All interviews were recorded with participant consent, and Microsoft Teams provided automatic transcription. In one case where the automatic transcription with Teams failed, a third-party transcription tool from the recording was utilized to maintain consistency in data capture.

Before each interview, participants were explicitly asked for consent regarding recording and transcription. They were assured that any sensitive information would be handled confidentially and that their insights would only be reported in an aggregated form to maintain anonymity. All participants provided verbal consent to these conditions.

The final sample represented a diverse cross-section of industries and geographical locations. Participants were based in Germany, Austria, and the Netherlands, providing a Central European perspective on AI adoption. The industry distribution included consulting firms, the medical industry, the sustainability sector, and technology companies, offering insights into AI implementation across various organizational contexts. The participants, all male and aged between 23 and 39, held diverse educational qualifications ranging from Bachelor's to Master's degrees and PhDs. They worked in organizations of varying sizes, from small consultancies with approximately 10 employees to global technology and consulting companies with a 6-digit employee headcount.

Initial contact with potential participants was made 1-2 weeks before the interviews. Interview questions were not shared beforehand to maintain spontaneity and avoid prepared responses. While no monetary compensation was offered for participation, participants were promised access to the final research report. All interviews were conducted in English to ensure data collection and analysis consistency.

\end{multicols}

\begin{figure}[!ht]
    \centering
    {\includegraphics[width=0.9\linewidth]{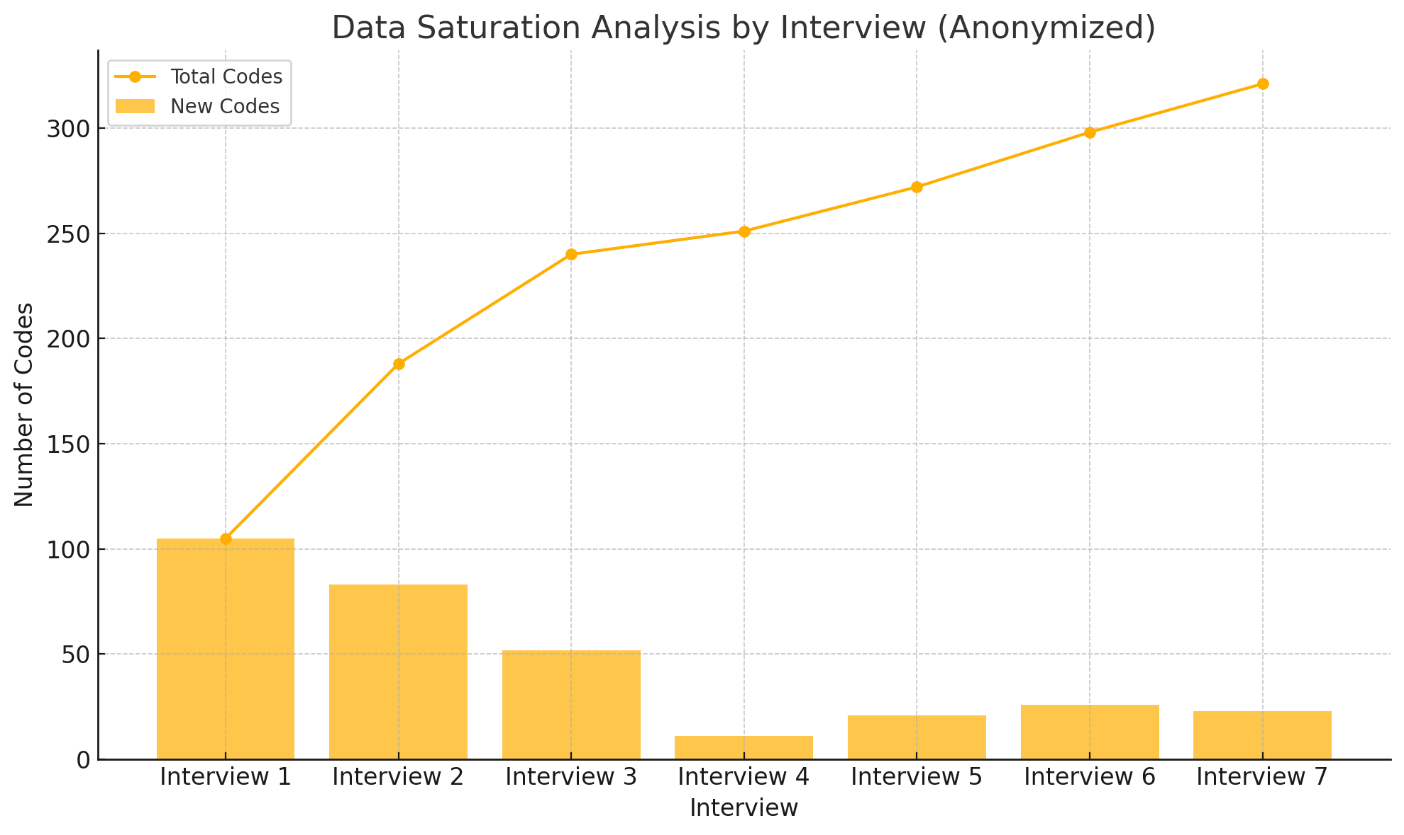}}
    \caption{Data Saturation Analysis}
    \label{fig:fig2}
\end{figure}

\begin{multicols}{2}

\subsection{Data Analysis}

Guided by a constructivist epistemology and employing the Gioia method \citep{Gioia2013} within a grounded theory framework \citep{Charmaz2012}, the analysis aimed to develop an empirically grounded theoretical understanding of how individual perceptions of AI limitations shape organizational readiness for AI adoption. The data analysis involved iterative coding, constant comparison, and progressive abstraction from initial participant statements to higher-level theoretical constructs. This structured approach ensured that emerging insights remained closely tied to the data while allowing for generating a novel theoretical framework.

The Gioia methodology provides a systematic, inductive process for qualitative data analysis that integrates participants' views with more abstract theoretical concepts. The analysis progressed through three main stages: (1) First-Order (Open) Coding, (2) Second-Order (Axial) Coding, and (3) Aggregate Dimensions \citep{Gioia2013}.

This process was iterative and reflective. Throughout the analysis, I constantly compared new codes and themes against previously coded data, refining concepts and ensuring consistency. Large language models assisted in the data analysis process in a human-supervised way through their text understanding affordance.

\subsubsection{First-Order Coding}

All seven interviews were transcribed and reviewed in their entirety. The initial coding began by closely reading each transcript line-by-line and assigning codes that captured the meaning of participant statements. At this stage, the codes remained descriptive and "participant-centric," avoiding premature interpretation. For example, statements about employees experimenting with AI tools were coded with phrases like "direct experience is crucial" or "hands-on learning approach." Similarly, when participants discussed informal knowledge-sharing networks among peers, codes such as "peer-based experience sharing" and "informal networks" emerged.

This first iteration of coding yielded a broad set of approximately 320 unique first-order codes (see Appendix 12.3 for the Initial Code Catalogue). These codes reflected a wide range of experiences, including the formation of AI limitations awareness, how trust and skepticism arose from direct experiences, the role of champions within organizations, and the influence of external pressures such as competition or regulation. Example quotes from the interviews were included in the Initial Code Catalogue.

During this phase, data saturation was actively monitored. As detailed in Section 3.3 and Appendix 12.2, the count of new codes diminished sharply after the third interview, indicating that the core concepts were stabilizing. Subsequent interviews mainly confirmed and refined existing codes rather than introducing entirely new concepts. This data saturation suggested a solid empirical foundation for moving toward higher-level conceptualization.

\subsubsection{Second-Order Themes}

In the second analysis stage, the initial codes were examined for similarities, differences, and conceptual relationships. This step involved moving from the raw "informant terms" toward more abstract, theoretical "researcher terms." Codes that shared conceptual relatedness or addressed related phenomena were clustered to form second-order themes.

For instance, numerous first-order codes related to how employees learned about AI limitations - through trial-and-error, observing peers, hands-on prototyping, and receiving training - were synthesized into themes like "Hands-on, Experiential Learning and Prototyping" and "Understanding and Communicating AI Limitations." Similarly, multiple codes describing how trust emerged incrementally through small successes and peer endorsements coalesced into the theme "Trust Building Through Incremental Successes."

Another example involved integrating discussions around internal advocates, knowledge-sharing communities, and informal networks, resulting in themes such as "Social Influence, Peer Learning, and Informal Networks" and "Champion and Ambassador Models." As the analysis progressed, it became clear that these themes were not isolated but were interrelated, often bridging technical understanding, social dynamics, and organizational structures.

A total of 25 second-order themes were distilled (see Appendix 12.4). These themes encompassed key factors such as governance structures, top-down vs. bottom-up adoption tensions, the shift from initial hype to realistic implementation understanding, the importance of cross-functional collaboration, and the balancing act between innovation and practical utility.

\subsubsection{Aggregate Dimensions}

In the final step, the second-order themes were clustered into higher-order, aggregate dimensions that captured the holistic patterns and processes emerging from the data. The goal was to develop a coherent theoretical framework showing how individual interpretations of AI limitations collectively influence organizational readiness. This integrative step led to the identification of five aggregate dimensions (see Appendix 12.5): (1) "Individual Sensemaking Foundations," (2) "Social and Organizational Learning Mechanisms," (3) "Organizational Integration and Governance," (4) "Expectation Management and Trust Development" and (5) "Long-Term Adaptation and Value Realization."

\section{Results}

Figure \ref{fig:fig3} presents a data structure visualizing how first-order codes aggregate into second-order themes and finally coalesce into the five aggregate dimensions. This structured representation helps illustrate the "funnel" of abstraction, starting from the richness of participant experience and ending with a theoretical framework that explains how individual-level interpretations of AI limitations shape organizational readiness.

\end{multicols}
\newpage
\begin{figure}[!ht]
    \centering
    {\includegraphics[width=0.7\linewidth]{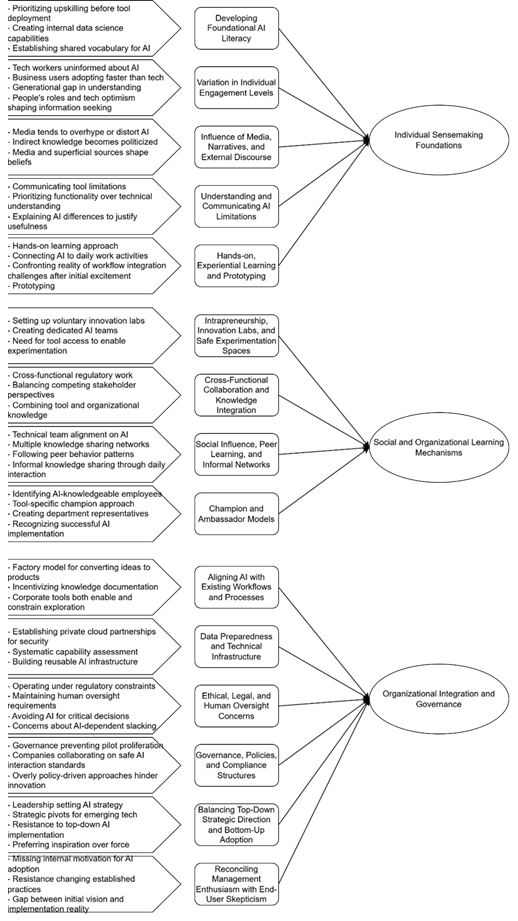}}
\end{figure}

\newpage
\begin{figure}[!ht]
    \centering
    {\includegraphics[width=0.8\linewidth]{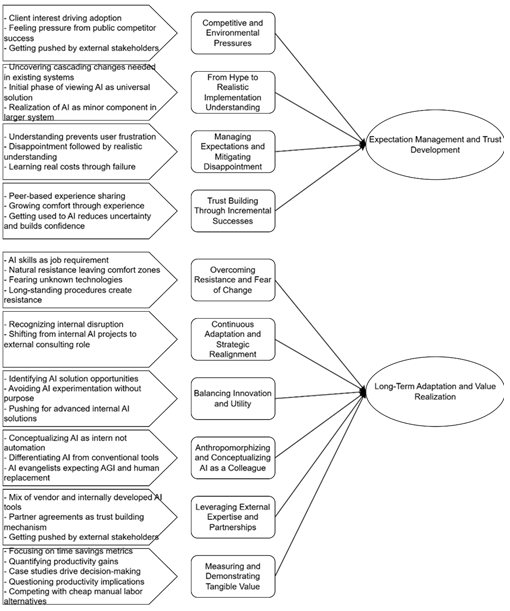}}
    \caption{Data Structure}
    \label{fig:fig3}
\end{figure}

\newpage
\begin{multicols}{2}

\section{Discussion}

\vspace{-3mm}
The empirical findings reveal complex interconnections between individual sensemaking processes and organizational AI readiness. Through analysis of the interview data, several key propositions emerged that help explain how organizations develop readiness for AI adoption. These propositions outline the pathways through which individual perceptions of AI limitations influence organizational readiness, mediated by processes of trust development, social learning, and organizational integration.

\end{multicols}

\begin{figure}[!ht]
    \centering
    {\includegraphics[width=1\linewidth]{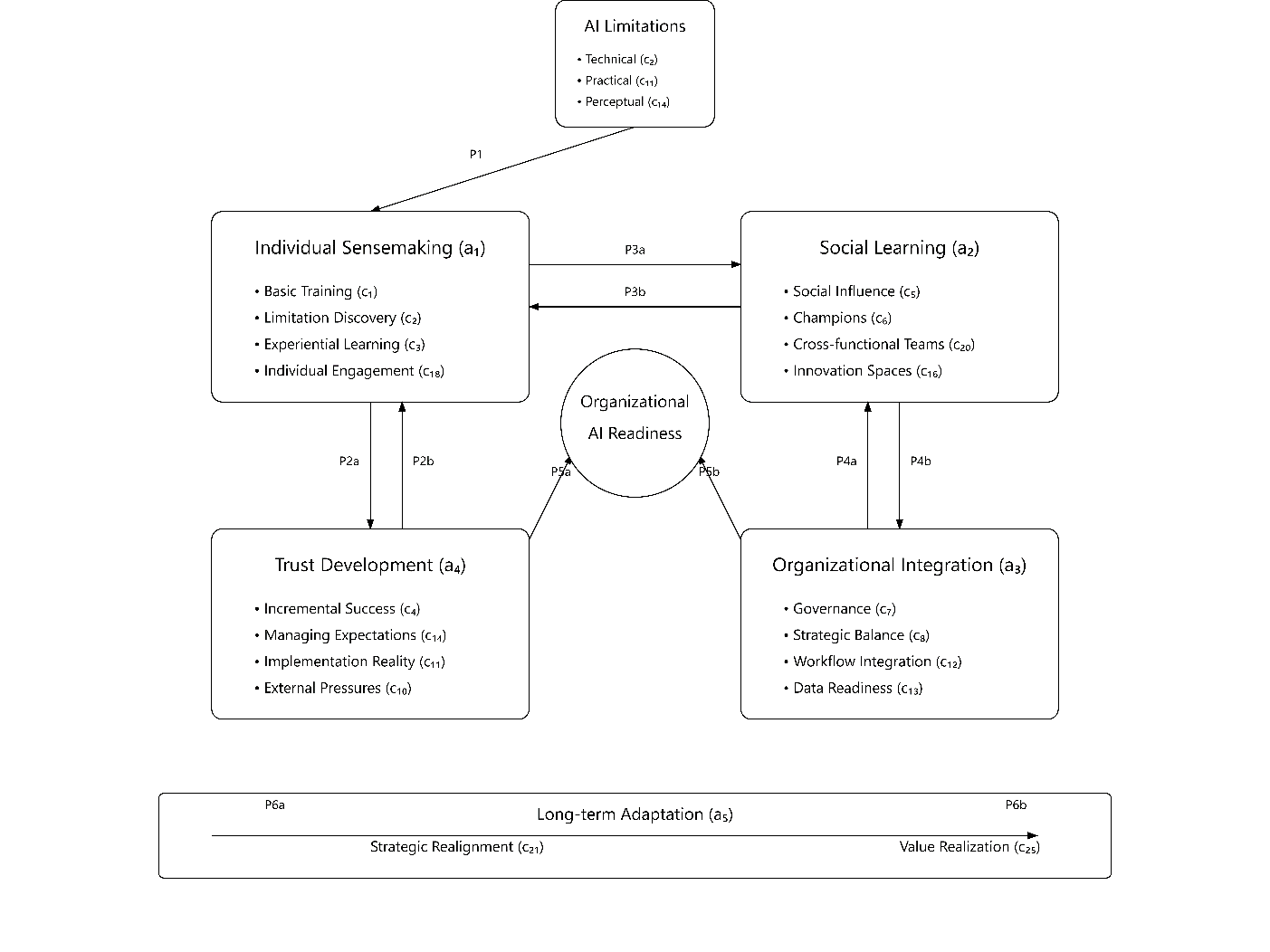}}
    \caption{Theoretical Model}
    \label{fig:fig4}
\end{figure}

\begin{multicols}{2}
    
\subsection{AI Limitations and Individual Sensemaking}
The analysis reveals how encounters with AI limitations trigger specific sensemaking processes that shape individual understanding and organizational readiness. This relationship manifests distinctly from the trust development and social learning mechanisms discussed earlier, focusing instead on the cognitive processes through which individuals interpret and internalize AI limitations. The data reveals how sensemaking evolves from reactive to proactive as individuals gain experience. Rather than just responding to interruptions, more experienced users actively extract and interpret cues about AI boundaries. As one participant explained, "If you have an idea of the limitation, you can [...] inform them about what you can do and what you can't do" (i\_238), indicating a shift from passive discovery to active extraction of cues about limitations. This transition resonates with \cite{Weick2005} observation that sensemaking often emerges from noticing and bracketing of ambiguous cues. In this context, AI limitations act as those cues; by recognizing and labelling them, individuals transform disruptions into actionable insights that reshape how they engage with AI.

This evolution aligns with \cite{Weick2005} description of sensemaking as an ongoing, retrospective process while extending it to show how limitation awareness enables more strategic technology engagement. The data suggests that individuals become better equipped to identify appropriate use cases as their interpretive frameworks become more sophisticated. One participant emphasized how "understanding how your company works [...] day-to-day operations [...] try to improve those" (i\_288), highlighting how refined interpretive frameworks enable more targeted implementation.

The data shows that specific technical constraints often serve as interruptions that trigger sensemaking processes. One participant described how "once I ran into an error in developing something, and I am like, OK, this is not possible" (i\_142), highlighting how technical barriers interrupt ongoing flows and prompt what \citep{Weick2005} describe as the noticing and bracketing of cues for interpretation. These confronting limitations represent clear instances where individuals must actively construct meaning from unexpected experiences.

This sensemaking process appears particularly potent when limitations interrupt existing assumptions. As one participant noted, "Someone showed me a prototype and was like, no, I made this. This is exactly what I couldn't do before" (i\_143), illustrating how limitation discoveries prompt active revision of understanding. This pattern of limitation discovery triggering interpretive processes suggests the first proposition:

\textit{P1: As individuals encounter and discover AI's limitations (e.g., hallucinations, token length constraints, biases), they are triggered to make sense of AI's capabilities and boundaries, refining their interpretations of how AI fits into their work.}

These insights advance theory by revealing how sensemaking about AI limitations differs from traditional technology sensemaking processes. Unlike conventional technologies, where limitations might be seen as constraints, AI limitations serve as interruptions that prompt ongoing cycles of noticing, interpretation, and action. This extends \cite{Weber2023} work on AI implementation capabilities by highlighting how individual sensemaking processes around limitations contribute to organizational capability development.

\subsection{Individual Perceptions and Trust Development}

The data reveals a relationship between how individuals make sense of AI limitations and their development of trust in AI systems. This relationship presents itself through two interrelated processes: the gradual development of trust through direct experience with AI limitations and the subsequent deepening of understanding through increased experimentation that this trust enables.

The findings show that individuals initially approach AI with varying skepticism and uncertainty, often influenced by media narratives and superficial coverage (c\_19). However, direct engagement with AI tools, mainly through experimentation and small-scale trials, reshapes these perceptions. One participant emphasized that "direct experience is crucial" (i\_38), highlighting how practical interaction helps individuals develop a better understanding of AI's capabilities and constraints. As \cite{Weick2005} argue, sensemaking is an ongoing process that is socially grounded and retrospective. In this case, experimentation with AI “talks” new experiences into being, which then become the basis for revising trust judgments. By looking back on both minor successes and failures, individuals develop more plausible and realistic expectations about AI’s utility and constraints.

The data highlights how understanding AI's limitations paradoxically builds rather than reduces trust. When individuals discover specific constraints - such as token length limits or potential for hallucinations - through controlled experimentation, they develop more realistic expectations about AI's role and capabilities. As noted by multiple participants, this understanding prevents the frustration and disappointment often resulting from inflated expectations (i\_42). This process resonates with \cite{Glikson2020} findings about how transparency about AI limitations can increase rather than undermine trust.  This pattern leads to the next proposition:

\textit{P2a: When people spend more time working with AI and understanding its limits (through training and experimenting), they trust it more, not less. This happens because they develop realistic expectations and see small successes.}

Furthermore, the findings indicate that once initial trust is established, it catalyzes a deeper phase of sensemaking and exploration. Participants described how successful experiences with AI encouraged them to push boundaries and explore new use cases. "Once you start using it [...] you're like, wow, it's actually way better" (i\_220), reflecting how positive experiences drive increased engagement. This created what interviewees observed as a recursive learning pattern: initial trust led to more experimentation, which generated a better understanding of capabilities and limitations.

The data shows that this deepened engagement manifests in more sophisticated testing of AI's boundaries. Participants described how growing trust made them "more open to explore or test new technologies" (i\_25) and enabled them to "check its credibilities, try and check where it works, where it doesn't work" (i\_28). This systematic probing of limitations represents a qualitative shift from initial cautious experimentation to more deliberate boundary testing. As noted by multiple participants, this deeper exploration often revealed nuanced limitations not apparent in initial use, such as specific contexts where AI might produce inconsistent results or require additional verification. These observations lead to another connected proposition:

\textit{P2b: Once people trust AI (e.g., after successful projects or recommendations from colleagues), they become more willing to experiment with it in new ways. This exploration helps them better understand what AI can and cannot do.}

Together, these propositions suggest a mutually reinforcing relationship between sensemaking about AI limitations and trust development at the individual level. Initial sensemaking builds trust through realistic expectation setting and small wins, while established trust enables deeper personal exploration that allows a better understanding of limitations. This integrated understanding of trust development and AI limitation discovery appears promotional for organizational readiness, as it enables a realistic assessment of AI's potential while providing the psychological safety needed for meaningful individual experimentation.

The findings extend research on AI adoption by illuminating how individuals develop trust in AI systems despite, or perhaps because of, their recognition of AI's limitations. Rather than limitations acting as barriers, the findings suggest that understanding limitations through hands-on experience and small successes helps users develop realistic mental models that enable effective use. The key appears to be approaching AI adoption not as a one-time acceptance decision but as an ongoing process where individuals gradually refine their understanding through direct engagement, balancing appreciation of AI's capabilities with a clear awareness of its constraints.

\subsection{Individual Sensemaking and Social Learning}

The data reveals how individual sensemaking processes interact dynamically with social learning mechanisms to shape organizational readiness for AI adoption. This relationship manifests through two key propositions highlighting how personal insights catalyze collective learning, reshaping individual understanding.

The data shows that as individuals develop more precise insights into AI through direct experience, they naturally share these discoveries with colleagues. Multiple participants described how employees who gained hands-on experience with AI became eager to share what they had learned. For instance, one participant noted that "colleague will start to see you that you are using chat [...] then you start to feel that I need that too" (i\_29), highlighting how individual discoveries spark interest in others. This sharing occurred through various channels - informal conversations, champion networks, and innovation labs (i\_151, i\_22, i\_229) - creating multiple pathways for knowledge dissemination. \cite{Weick2005} remind that sensemaking is inherently social; it builds on communication that talks events into existence and promotes shared understanding. In this study, employees’ peer-to-peer exchanges mirror that dynamic: individual discoveries become group insights through the ongoing construction and negotiation of meaning within these networks.

This pattern was particularly evident in how organizations leveraged "black belts"/ champions (i\_146, i\_24)  or department representatives who "volunteer to be representative of using AI tools and to share knowledge" (i\_147). Having developed personal understanding through experience, these individuals became missionaries for spreading practical insights about AI limitations and capabilities throughout their departments. Participants emphasized how this peer-to-peer knowledge sharing proved "much better than we like as technical people to communicate that to them" (i\_29), suggesting that social learning benefits from the authenticity of peer experiences.

The data also revealed that organizations actively cultivated these knowledge-sharing dynamics by creating "voluntary venture labs" (i\_151) and establishing regular sharing sessions where employees could "tell us how are you using it and send us a picture" (i\_152). These structured opportunities for sharing personal insights amplified the natural tendency for individual learning to spark collective understanding. This leads to the next proposition:

\textit{P3a: When employees figure out what AI is good and bad at through direct experience, they share these insights with their coworkers through formal and informal ways.}

The relationship between individual sensemaking and social learning appears to be bidirectional. The data showed that as social networks and communities exchanged AI experiences, individuals began reinterpreting their views based on others' experiences. Participants described how "understanding improves through peer interaction and experimentation" (i\_29), suggesting that exposure to others' experiences helps refine personal interpretations of AI.

This collective influence on individual sensemaking was particularly evident in how organizations leveraged "echo chambers" (i\_162) and interconnected groups where it is "easy to bring this message" (i\_164). While the term "echo chambers" might carry negative connotations, in this context, it describes how shared experiences within departments or teams helped reinforce and refine individual understanding. The data showed that these social dynamics were successful when they included concrete examples, with participants noting how "trust builds through peer experience sharing" (i\_29).

The influence of collective experience on individual sensemaking aligns with Weick's (1995) emphasis on the social nature of sensemaking processes. The data revealed how "informal network narratives shape organizational readiness" (i\_29), suggesting that individual interpretations of AI are continuously refined through exposure to colleagues' experiences, successes, and failures. This recursive relationship between individual and collective understanding leads to proposition P3b:

\textit{P3b: As people share their AI experiences in teams and networks, individuals update their understanding of AI based on their colleagues' successes, failures, and best practices. This creates a cycle where personal and group learning reinforce each other.}

These propositions advance theory by revealing how AI adoption catalyzes unique dynamics between individual sensemaking and collective learning processes. While sensemaking theory has traditionally focused on how individuals interpret novel technologies \cite{Weick1995}, and organizational learning frameworks examine knowledge transfer across levels \citep{Crossan1999}, the findings show that AI's distinctive characteristics - its opacity, evolving capabilities, and context-dependent performance - require continuous interplay between personal discovery and social validation. The reciprocal relationship between individual exploration and collective sensemaking appears especially critical for AI adoption, as the technology's complexities mean that no single person's understanding is sufficient; instead, organizational readiness emerges through the ongoing synthesis of diverse individual experiences shared through social networks.

\subsection{Social Learning and Integration}

Building on the previous findings about trust development and social learning, the data reveals distinct patterns in how formalized structures and social learning mechanisms interact to shape organizational AI readiness. This relationship emerges through two key processes: how mature integration frameworks enable systematic knowledge sharing and how collective learning drives organizational adaptation.

The data shows successful organizations develop specific structural mechanisms to facilitate knowledge exchange. Participants described the creation of "voluntary venture labs" (i\_151) and regular sharing sessions where employees would "tell us how are you using it and send us a picture" (i\_152). These structured opportunities moved beyond informal conversations to create dedicated spaces for knowledge exchange. From \cite{Weick2005} viewpoint, these formalized routines exemplify how organizations can actively shape the environment in which sensemaking unfolds. By creating explicit forums—such as “voluntary venture labs” or cross-functional teams—organizations enact structures that channel how cues are noticed, interpreted, and retained over time. One participant emphasized how organizations established "domain-specific AI communities" (i\_65) focused on particular business functions like "AI for tax, AI for others, AI for assurance, AI for knowledge management" (i\_65), indicating how formal structures enabled targeted knowledge sharing.

Multiple participants highlighted how these formal mechanisms helped bridge departmental boundaries. One noted that "capabilities and capacity spread across departments that somehow have to communicate" (i\_64), while another described "cross-departmental knowledge sharing emerges" as a key outcome of structured integration efforts. The importance of formal support was particularly evident in observations about "creating presentation opportunities" (i\_64) and "encouraging external knowledge sharing" (i\_64). This alignment of structure and learning leads to the next proposition:

\textit{P4a: As formal AI integration efforts (e.g., governance committees, embedded AI workflows, robust data infrastructure) mature, they enable more structured environments - like cross-functional teams - that improve social learning and knowledge exchange among employees.}

Furthermore, the findings indicate that collective learning drives specific organizational changes. Participants described how shared experiences led organizations to formalize "innovation process[es] from idea" (i\_95) and establish systematic approaches to pilot projects. One participant noted how "collective learning through reporting" (i\_29) helped organizations identify patterns and standardize successful approaches. This pattern was particularly evident in how "organizations build internal capabilities through testing" (i\_27), suggesting that shared learning experiences inform formal capability development.

The data reveals that this process involves creating new organizational roles and structures. Multiple participants described the emergence of "specialized teams in response to AI hype" (i\_177) and efforts at "coordinating across departments" (i\_178). One participant emphasized how organizations began creating dedicated AI teams (i\_298) based on collective learning about what worked. These observations lead to proposition P3b:

\textit{P4b: When teams share their AI success stories and best practices, organizations are more likely to make AI a permanent part of their operations by updating their processes and systems.}

These propositions illuminate how social learning and formal integration mechanisms reinforce each other. The findings extend research on organizational learning by showing how AI adoption requires both structured knowledge-sharing environments and the ability to translate collective insights into formal organizational changes. Unlike previous technologies, AI's complexity and evolving nature demand continuous interplay between social learning and structural adaptation. This analysis suggests that organizations can actively facilitate this virtuous cycle by creating formal spaces for knowledge exchange while remaining flexible enough to incorporate emerging insights into their structures and processes. The key appears to balance structured support for learning with the ability to evolve organizational frameworks based on collective experience.

\subsection{Trust and Integration Effects on Organizational Readiness}

The data reveals two distinct but interlinked pathways through which organizations develop readiness for AI adoption. First, trust development emerges from individual sensemaking processes (Section 5.1, 5.2) and catalyzes broader AI acceptance. Second, organizational integration mechanisms - spawned by social learning (Section 5.3, 5.4) - shape the structural and procedural capabilities that sustain AI adoption. Together, these two factors (trust and integration) help translate micro-level learning (individual sensemaking and social learning) into macro-level change (organizational readiness).

Trust in AI repeatedly surfaces in the findings as a outcome of individual-level sensemaking. Individuals experimenting with AI and discovering its limitations (e.g., hallucinations, token-length constraints) refine their mental models of what AI can and cannot do (i\_14, i\_238). This process of hands-on sensemaking (i\_38, i\_142) corrects inflated expectations (i\_42), develops realistic understanding, and builds confidence in AI (i\_56). In turn:

\textit{P5a: Trust translates individuals' refined sensemaking of AI into broader organizational readiness. As trust in AI becomes widespread, the organization becomes more willing to undertake new AI initiatives.}

The role of trust is evident in how individuals who "spend more time working with AI and understanding its limits" (P1a) develop not only accurate expectations but also a psychological readiness to champion AI within their teams (i\_29). Positive experiences - "wow, it's actually way better" (i\_220) - fuel further exploration, demonstrating how trust channels individual insight into deeper organizational buy-in. Even imperfect outcomes can "still increase trust" (i\_265) if users come away with a clearer sense of AI's potential and boundaries. Consequently, trust development - anchored in realistic appraisals of AI - reduces perceived risk, mitigates resistance, and paves the way for more extensive AI adoption across the organization (i\_175, i\_199).

Parallel to trust, organizational integration emerges from collective social learning processes. As people share experiences and best practices (Sections 7.2, 7.3), organizations gradually create formal mechanisms - for example, "domain-specific AI communities" (i\_65), "voluntary venture labs" (i\_151), or "specialized teams in response to AI hype" (i\_177). These structures allow governance, cross-departmental communication channels, and shared technical infrastructure, ensuring that early lessons do not remain isolated within pockets of the organization (i\_64, i\_72).

\textit{P5b: Organizational integration describes the relationship between social learning and organizational readiness. As AI becomes embedded in formal structures and processes, the organization develops more substantial capabilities and frameworks for future AI adoption.}

This is reflected in how shared successes and "domain-specific" case studies (i\_29, i\_96) get systematized into policies and workflows (i\_31), establishing consistent practices for AI governance, risk assessment, and capability building. Organizational integration thus institutionalizes the collective insights gained via peer-to-peer knowledge exchange and champion networks (i\_29, i\_146–i\_149). Such integration ensures that insights from social learning get translated into formal support structures, ultimately accelerating AI adoption speed, scale, and sustainability.

\subsection{Long-term Adaptation}

Finally, a temporal dimension of AI adoption emerged in the data, revealing how organizations' approach to AI implementation and value realization evolves over extended periods. This longitudinal perspective provides insights into how initial experiences shape long-term adaptation strategies and eventually influence organizational readiness for AI.

The data shows that organizations' relationship with AI technologies undergoes maturation phases. Initially, as one participant described, organizations often move "from overconfidence to disappointment" (i\_16) before developing more nuanced approaches. This evolution was not linear but involved iterative cycles of learning and adaptation. Another participant emphasized how continuous experimentation led to deeper understanding: "Once you start using it [...] you're like, wow, it's actually way better" (i\_220), highlighting how direct experience shapes organizational approaches over time.

The findings reveal that successful organizations demonstrated an ability to learn from both positive and negative experiences, using these insights to refine their implementation strategies. One participant noted that "even if the model becomes worse than the expectations, it still increases trust" (i\_265), suggesting that even unsuccessful implementations contributed to organizational learning. This pattern was particularly evident in how organizations adjusted their resource allocation and structural arrangements over time. For instance, one participant described how their organization "sourced the three people that were the most affiliated with AI developments into a special task force" (i\_177) as a response to accumulated implementation experience.

This approach was not merely reactive but became increasingly strategic as organizations gained experience. One participant emphasized how "understanding how your company works [...] day-to-day operations [...] try to improve those" (i\_288) became central to their approach over time. The data shows that organizations that successfully sustained AI adoption developed systematic approaches to capturing and applying lessons learned. Another participant noted that the "continuous mixing of experts in […] different projects" (i\_72) enabled ongoing knowledge transfer and capability development. These observations about organizational learning and adaptation over time lead to the first proposition:

\textit{P6a: As organizations gain more experience with AI - successes and failures - they adjust their strategies, resources, and structures to keep AI aligned with their long-term goals.}

The data further reveals how sustained engagement with AI leads to evolving patterns of value realization. Initially, organizations often focused on immediate efficiency gains but developed more sophisticated approaches to value creation over time. One participant observed that "organizations build internal capabilities through testing" (i\_27), suggesting a gradual capability development process. This evolution was particularly evident in how organizations moved from isolated AI experiments to more integrated approaches.

The temporal aspect of value realization emerged strongly in how organizations learned to leverage AI effectively. One participant described how "experience leads to belief updating" (i\_144), indicating an iterative learning and value discovery process. Another noted that their organization needed to "make a real process out of it and tell others [...] use this process as it's time efficient" (i\_183), showing how initial successes were systematized into repeatable approaches.

This pattern of evolving value realization aligns with \cite{Weber2023} findings about organizations developing specific implementation capabilities over time, including AI project planning, co-development, data management, and model lifecycle management. However, the data extends this understanding by showing how value realization becomes increasingly sophisticated as organizations gain experience. As one participant explained, "If you have an idea of the limitation, you can [...] inform them about what you can do and what you can't do" (i\_238), suggesting that accumulated experience enables more strategic deployment of AI capabilities. The long-term pattern of value realization and its impact on organizational embedding leads to the second proposition:

\textit{P6b: Over time, continuous learning and integration of AI leads to real organizational benefits, which encourages organizations to embed AI more deeply into their culture and operations.}

These propositions advance our understanding of organizational readiness by highlighting its temporal dimension. While previous research has often treated readiness as a static state, the findings suggest it is better understood as an evolving capability that develops through cycles of learning and adaptation. This builds on \cite{Johnk2021} framework but adds specific insights about how organizations' ability to realize value from AI improves over time.

The findings reveal that successful long-term adaptation requires organizations to maintain flexible structures while building systematic approaches to learning. One participant noted that they needed to "create groups to exchange knowledge" (i\_29) while establishing formal processes for capturing and applying lessons learned. This balance between flexibility and systematization emerged as crucial for sustained AI adoption.

This temporal perspective illuminates how organizations move beyond viewing AI adoption as a discrete change initiative to see it as an ongoing process of organizational evolution. While this aligns with \cite{Crossan1999} organizational learning framework, it adds specific insights into how organizations learn to work with AI technologies over time. The data suggests that successful organizations develop the ability to continuously evolve their approach to AI based on accumulated experience while maintaining alignment with strategic goals.

The findings extend previous research by showing how initial experiences with AI shape longer-term organizational responses. Organizations that successfully sustained AI adoption demonstrated an ability to learn from successes and failures, using these insights to develop more sophisticated approaches to implementation over time. This temporal dimension suggests that organizational readiness for AI is not achieved at a single point but continues to evolve as organizations gain experience and develop a more nuanced understanding of creating value through AI technologies. Consistent with \cite{Weick2005}, this readiness emerges through ongoing sensemaking processes, where individuals interpret AI experiences retrospectively, construct trust through realistic expectations, and share knowledge socially. Over time, iterative learning cycles refine practices, align trust with integration, and enable organizations to adapt to the unique challenges of AI, demonstrating that readiness is a dynamic, evolving capability rather than a static state.

\section{Practical Implications}

The findings of this study indicate that organizational readiness for AI adoption depends not merely on technical infrastructure and leadership directives but also on how employees form and diffuse their understanding of AI's limitations. A first managerial implication is that organizations should prioritize building foundational AI literacy through hands-on experimentation. This aligns with \cite{Henry2022} findings on the importance of human-machine teaming and experiential learning. Conceptual overviews, though important, are insufficient for helping individuals understand AI's actual boundaries, such as hallucination tendencies or token-length constraints. When employees are encouraged to test AI tools in sandbox environments or pilot projects, they develop more realistic expectations and cultivate a measured confidence in the technology's potential, consistent with \cite{Weber2023} emphasis on implementation capabilities. These incremental "wins" help mitigate the disillusionment that often arises when inflated expectations clash with technical realities. Organizations can actively facilitate this sensemaking by creating formal and informal knowledge-sharing spaces, cultivating champion networks, and encouraging cross-functional exchange, as supported by \cite{Kelley2022} identification of success factors. However, the data emphasizes that these social learning mechanisms work best when they emerge organically from genuine individual insights rather than top-down initiatives.

A second implication underscores the significance of cultivating trust gradually through tangible success stories and demonstrable improvements in workplace tasks, aligning with \cite{Glikson2020} findings on trust development in AI systems. Even if AI systems produce imperfect outcomes, employees who see clear efficiency gains become more open to more advanced experimentation. This trust-building process benefits significantly from the support of champions and peer-to-peer advocacy, reflecting \cite{Siau2018} emphasis on the importance of expert endorsements and iterative user interactions in increasing trust. Formal directives from senior management can initiate AI adoption. However, as \cite{Felemban2024} demonstrate, genuine, sustained acceptance often flows from informal networks where colleagues coach one another and share pragmatic guidance.

Beyond these points, the study highlights that structured governance and clear policies are necessary to manage potential risks without stifling innovation, consistent with \cite{Johnk2021} framework of core readiness domains. Particularly in regulated sectors, compliance, and liability concerns demand a thoughtful approach that offers employees enough flexibility to discover practical uses for AI while maintaining safeguards around data privacy and ethical standards. However, such oversight remains most effective when paired with a robust internal communication strategy that reconciles the enthusiasm of senior leaders with the daily realities and valid concerns of end-users, as emphasized by \cite{MikelHong2024} regarding the role of leadership in addressing resistance.

A further implication pertains to the need for continuous alignment between AI initiatives and broader organizational processes and culture. Rather than treating AI as a standalone innovation, managers can embed it into existing operations and strategic roadmaps, as suggested by \cite{Heimberger2024} findings on process integration. Sustained readiness similarly relies on adapting as technology evolves, so policies, pilot projects, and success metrics must be reviewed continuously to capture new opportunities or address emerging limitations. Over time, demonstrating tangible outcomes strengthens organizational buy-in. It justifies further investment in AI's technical and human dimensions, aligning with \cite{Yang2024} observations about how organizations perceive and realize AI's affordances.

\section{Limitations and Future Directions}

While this research draws strength from its qualitative, expert-interview approach and provides in-depth perspectives on individual sensemaking, three main limitations warrant attention and suggest avenues for future inquiry.

First, the study focuses on individual-level interpretations, offering limited insight into how these perceptions coalesce into collective readiness across organizational tiers \citep{Crossan1999}. Given the intermediate state of theory connecting individual perceptions to organizational readiness, future research could employ hybrid methods combining qualitative and quantitative approaches \citep{Edmondson2007} through longitudinal or multi-level case studies. Such nested designs involving frontline employees and middle managers could investigate how individual sensemaking about AI's constraints ultimately drives or hinders systemic transformation \citep{Maitlis2014,Orlikowski1994,Weiner2009}.

Second, the purely qualitative design, while appropriate for exploring novel phenomena \cite{Edmondson2007}, foregrounds subjective narratives, which raises concerns about potential bias and the risk of overemphasizing anecdotal success stories or attributing failures to external factors. As theory in this domain matures, scholars could integrate mixed methods approaches, coupling in-depth interviews with large-scale quantitative surveys or archival data to validate and expand upon the themes identified here. For instance, measuring constructs such as trust, readiness, or perceived limitations at scale would help verify whether the patterns observed in interviews generalize to broader organizational contexts.

Third, the small sample of AI-focused experts, while rich in detail and appropriate for nascent theory development \citep{Edmondson2007}, may tilt findings toward those who are technologically forward-thinking or predisposed toward AI experimentation. Future work could involve a more diverse cohort of informants—such as frontline staff, middle managers, or external partners—and broaden the industry scope to gauge whether these insights remain consistent across different sectors. This expanded approach can reveal the extent to which varying organizational cultures, regulatory environments, or leadership styles shape readiness and adoption trajectories.

Finally, given the dynamic AI landscape, ongoing advances in model architectures, data processing, and training methods may mitigate or eliminate some limitations identified here \citep{Bommasani2022,Bubeck2023}. As theory development progresses from nascent to intermediate stages, longitudinal research that examines AI adoption over extended periods could illustrate how early, significant barriers diminish once organizations refine data pipelines, cultivate new skill sets, or implement governance structures. Such designs would clarify how shifting technical and organizational landscapes influence the evolution of trust, readiness, and overall AI strategy \citep{Henry2022,Weick2005,Crossan1999}.

\section{Conclusion}

This work examines how individual perceptions of AI limitations influence organizational readiness for AI adoption. The findings reveal a dynamic interplay between individual sensemaking processes, social learning mechanisms, and formal organizational structures. When employees encounter AI limitations through hands-on experience, they develop more realistic expectations and greater trust in the technology, mainly when supported by peer networks and champion systems. Organizations that successfully translate these individual and collective insights into formal governance structures and processes are better positioned for sustainable AI adoption. The research demonstrates that organizational AI readiness is not a static state but an evolving capability that emerges through the continuous interaction between individual understanding, social learning, and organizational adaptation. This suggests that organizations should approach AI adoption not as a one-time implementation but as an ongoing strategic learning process that balances innovation with practical constraints.

\bibliographystyle{apalike}
\bibliography{references}
\newpage
\end{multicols}
\section{Appendix}
\subsection*{Appendix 1: Interview Guide}
\textbf{Introduction}

\begin{itemize}
    \item 	Welcome and Introduction
\item	Purpose of the Study
\begin{itemize}
\item The purpose of this study is to gain insights from experts like you on the factors that impact AI adoption in organizations, specifically focusing on perceptions of AI limitations.
\end{itemize}
\item	Confidentiality
\item	Consent to recording

\end{itemize}

\textbf{Context}
\begin{itemize}
\item	Could you briefly describe your experience with AI implementation projects in organizations?
\end{itemize}
\textbf{Theme 1: Perception Formation of AI Limitation}

\begin{itemize}
\item	Based on your observations, how do individuals in organizations typically develop their understanding of AI limitations?

\begin{itemize}
    \item 	\textit{Probe: Role of professional background (technical vs non-technical?)}
\item	\textit{Probe: Impact of direct experiences vs. indirect knowledge}
\item	\textit{Probe: Through formal training, peer discussions, hands-on experience, or other means?}
\item	\textit{Probe: Industry context influence}
\item	\textit{Probe: External discourse influence (E.g. media coverage? Failed projects? Successful projects in other organizations?)}
\end{itemize}

\end{itemize}

\textbf{Theme 2: Individual and Collective Sensemaking}

\begin{itemize} 
\item 	How have you seen individuals interpret and make sense of their experiences with AI limitations?

\begin{itemize}
    \item 	\textit{Probe: Trust-building}
\item	\textit{Probe: Contradictions between their expectations and actual AI performance}
\item	\textit{Probe: Role of professional identity}
\item	\textit{Probe: Impact of past experiences}
\item	\textit{Probe: Experimentation}
\item	\textit{Probe: Process of updating interpretations (What triggers change over time?)}

\end{itemize}

\item 	In your experience, how do collective interpretations of AI limitations develop within organizations?

\begin{itemize}
    \item 	\textit{Probe: Knowledge sharing mechanisms (How do organizational stories or narratives about AI successes/failures spread?)}
\item	\textit{Probe: Informal networks' role}
\item	\textit{Probe: Leadership influences}
\item	\textit{Probe: Resolution of conflicting perspectives}

\end{itemize}

\end{itemize}

\textbf{Theme 3: Impact on Organizational Readiness}

\begin{itemize}
    \item 	How do individual understandings of AI limitations shape an organization's readiness for adoption?
\begin{itemize}
    \item 	\textit{Probe: Change management}
\item	\textit{Probe: Communication}
\item	\textit{Probe: Change in culture}
\item	Probe: Risk assessment
\item	\textit{Probe Cross-functional coordination}
\item	\textit{Probe: Changes in strategic planning}
\item	\textit{Probe: Adjustments to implementation timelines}
\item	\textit{Probe: Development of support structures}

\end{itemize}

\item How does an organization adapt its practices when confronted with AI limitations identified by employees? 

\begin{itemize}
    \item 	\textit{Probe: Resource reallocation  (Hiring? Skill development?)}
\item	\textit{Probe: Training program modifications}
\item	\textit{Probe: Process redesign efforts}

\end{itemize}

    \item 	What role do collective experiences with AI limitations play in shaping organizational capabilities? 

    \begin{itemize}
        \item 	\textit{Probe: Changes in collaboration patterns}
\item	\textit{Probe: Changes in decision-making processes}

    \end{itemize}
\end{itemize}

\textbf{Closing}

\begin{itemize}
    \item 	Based on your observations, what do you see as the most critical connection between individual perceptions of AI limitations and organizational readiness?
\item	Is there anything else you'd like to add about how perceptions of AI limitations influence organizational readiness for AI adoption?

\end{itemize}

Thanking the participant

\subsection*{Appendix 2: Data Saturation}

\begin{lstlisting}[language=JSON]

import pandas as pd

# Load the uploaded CSV file
file_path = "/mnt/data/Initial Codes - Sheet4.csv"
data = pd.read_csv(file_path)

# Display the first few rows of the dataset to understand its structure
data.head(), data.columns

# List of interviewees in the order they were conducted
interviewees = [
    "Interviewee 1",
    "Interviewee 2",
    "Interviewee 3",
    "Interviewee 4",
    "Interviewee 5",
    "Interviewee 6",
    "Interviewee 7",
]

# Extract the interviewee's name from each quote and create a new column for clarity
data['Interviewee'] = data['Quotes'].apply(
    lambda quote: next((name for name in interviewees if name in quote), None)
)

# Count occurrences of interviewees
interview_counts = data['Interviewee'].value_counts()

# Identify new codes introduced by each interviewee
codes_per_interviewee = data.groupby('Interviewee')['Code'].apply(lambda codes: codes.unique())

# Check cumulative data saturation (unique codes found per interviewee as they appear in order)
saturation_data = {}
unique_codes = set()

for interviewee in interviewees:
    if interviewee in codes_per_interviewee:
        new_codes = set(codes_per_interviewee[interviewee]) - unique_codes
        unique_codes.update(new_codes)
        saturation_data[interviewee] = {"New Codes": len(new_codes), "Total Codes": len(unique_codes)}

# Convert saturation data to a DataFrame for better visualization
saturation_df = pd.DataFrame.from_dict(saturation_data, orient='index')
import ace_tools as tools; tools.display_dataframe_to_user(name="Interview Data Saturation Analysis", dataframe=saturation_df)

import matplotlib.pyplot as plt

# Create a graph for data saturation
plt.figure(figsize=(10, 6))

# Plot total codes and new codes
plt.plot(
    saturation_df.index, saturation_df["Total Codes"], marker="o", label="Total Codes"
)
plt.bar(saturation_df.index, saturation_df["New Codes"], alpha=0.7, label="New Codes")

# Add titles and labels
plt.title("Data Saturation Analysis by Interview", fontsize=16)
plt.xlabel("Interviewee", fontsize=12)
plt.ylabel("Number of Codes", fontsize=12)
plt.legend()
plt.grid(axis="y", linestyle="--", alpha=0.7)

# Show the graph
plt.tight_layout()
plt.show()


\end{lstlisting}

\subsection*{Appendix 3:	Initial Code Catalogue}

\renewcommand{\arraystretch}{1.5}
\begin{longtable}
{|l|>{\raggedright\arraybackslash}p{100pt}|>{\raggedright\arraybackslash}p{290pt}|}\hline
\textbf{Identifier}&	\textbf{Code}&	\textbf{Representative Quotes}\\ \hline

i\_1   & dual role as internal solution   provider and employee support                      & "{[}...{]} we developed AI   solutions for {[}pseudonymized entity{]} itself {[}...{]} we support {[}pseudonymized   entity{]} employees"                                                                                                                                                                                                                                                                                                                                                                                                                                                                                                                                                                                                                                                                                                                                                                                                                                                                                                                                                                                                                                                                                                                                                                                                                                                                                                                                                                                                                                                     \\
\hline

\end{longtable}

\subsection*{Appendix 4:	Second-order Concept Catalogue}
\renewcommand{\arraystretch}{1.5}
\begin{longtable}{|l|>{\raggedright\arraybackslash}p{100pt}|>{\raggedright\arraybackslash}p{100pt}|>{\raggedright\arraybackslash}p{195pt}|} \hline
    Code&	Initial Codes	&Theme&	Description\\ \hline

c\_1  & i\_4, i\_9, i\_10, i\_11, i\_12, i\_13,   i\_17, i\_18, i\_50, i\_101, i\_174, i\_242, i\_259, i\_288 & Developing Foundational AI   Literacy                                & Organizations foster basic   understanding of AI capabilities, terminology, and concepts to prepare   employees for meaningful tool usage.                                                            \\ \hline
c\_2  & i\_14, i\_16, i\_59, i\_61, i\_148,   i\_196, i\_238, i\_246, i\_63, i\_142, i\_306, i\_314, i\_318   & Understanding and Communicating AI   Limitations                     & Employees must learn where AI   falls short—hallucinations, token length limits, biases—to form realistic   expectations, avoid frustration, and refine use cases.                                    \\ \hline
c\_3  & i\_38, i\_49, i\_54, i\_130, i\_173,   i\_190, i\_217, i\_219, i\_220, i\_264, i\_265, i\_289         & Hands-on, Experiential Learning   and Prototyping                    & Direct experience—from small tests   to prototypes—helps individuals understand AI's capabilities, revealing   hidden constraints and building accurate mental models.                                \\ \hline
c\_4  & i\_29, i\_56, i\_57, i\_105, i\_173,   i\_175, i\_209, i\_220, i\_264, i\_265, i\_240, i\_308         & Trust Building Through Incremental   Successes                       & Trust grows gradually as users see   AI deliver small, reliable benefits. Positive peer experiences, quick wins,   and iterative improvements cultivate confidence.                                   \\
\hline 

\end{longtable}

\subsection*{Appendix 5: Aggregate Dimensions}

\renewcommand{\arraystretch}{1.5}
\begin{longtable}
{|p{90pt}|p{100pt}|p{240pt}|}\hline
\textbf{Aggregate Dimension (a\_x)}&	\textbf{Second-Order Themes (c\_x) Integrated}&	\textbf{Description of Aggregate Dimension}\\ \hline

a\_1: Individual Sensemaking   Foundations            & c\_1: Developing Foundational AI Literacy\newline    c\_2: Understanding and Communicating AI Limitations\newline    c\_3: Hands-on, Experiential Learning and Prototyping\newline    c\_18: Variation in Individual Engagement Levels\newline    c\_19: Influence of Media, Narratives, and External Discourse                                                                                                  & Individuals develop initial   perceptions of AI capabilities and limitations through basic education,   direct experimentation, and external narratives. Engagement levels differ,   and media influences shape early assumptions, forming the groundwork for   organizational readiness.                                                                     \newline \\ \hline
a\_2: Social and Organizational   Learning Mechanisms & c\_5: Social Influence, Peer   Learning, and Informal Networks\newline    c\_6: Champion and Ambassador Models\newline    c\_20: Cross-Functional Collaboration and Knowledge Integration\newline    c\_16: Intrapreneurship, Innovation Labs, and Safe Experimentation Spaces                                                                                                                             & Social processes, such as peer   learning, champions, cross-functional teams, and safe innovation spaces,   transform individual knowledge into collective competencies. These mechanisms   encourage shared understanding, alignment, and the scaling of AI adoption   throughout the organization.                                                          \newline \\
\hline

\end{longtable}

\end{document}